\documentclass[preprint, 3p, twocolumn]{elsarticle}
\usepackage{mathrsfs}
\usepackage{amsmath}
\usepackage{amssymb}
\usepackage{graphicx}
\usepackage{subfigure}
\usepackage{amssymb}
\usepackage{amsthm}
\pdfoutput=1
\usepackage{pgf}
\usepackage{pdfpages}
\DeclareGraphicsExtensions{.png}
\usepackage{lineno}
\usepackage{comment}

\journal{Nuclear Instruments and Methods A}

\begin{document}

\begin{frontmatter}

\title{Numerical simulation of charging up, accumulation of space charge and formation of discharges}

\author[A]{Purba Bhattacharya}
\ead{purba.bhattacharya85@gmail.com}
\author[B]{Promita Roy}
\author[C]{Tanay Dey}
\author[D]{Jaydeep Datta}
\author[E]{Prasant K. Rout}
\author[F]{Nayana Majumdar}
\author[G]{Supratik Mukhopadhyay}
\address[A]{Department of Physics, School of Basic and Applied Sciences, Adamas University, Kolkata, Inda}
\address[B]{Department of Physics, , Virginia Polytechnic Institute \& State University, VA, United States}
\address[C]{School of Physical Sciences, National Institute for Science Education and Research, Jatni, Khurda, Odisha, India}
\address[D]{Center for Frontiers in Nuclear Science, Department of Physics and Astronomy, Stony Brook University, 100 Nicolls Road, Stony Brook, New York, 11794, USA}
\address[E]{National Central University, Taoyuan City, Taiwan}
\address[F]{Applied Nuclear Physics Division, Saha Institute of Nuclear Physics, Kolkata, India}
\address[G]{Retired from Applied Nuclear Physics Division, Saha Institute of Nuclear Physics, Kolkata, India}

\begin{abstract}
Aging and stability of gaseous ionization detectors are intricately related to charging up, accumulation of space charge and formation of discharges.
All these phenomena, in their turn, depend on the dynamics of charged particles within the device. Because of the large number of particles involved and their complex interactions, the dynamic processes of generation and loss of charged particles, and their transport within the detector volume are extremely expensive to simulate numerically. In this work, we propose and evaluate possible algorithms / approaches that show some promise in relation to the above-mentioned problems.
Several important ionization detectors having parallel plate configurations, such as GEM, Micromegas, RPCs and THGEMs, are considered for this purpose.
Information related to primary ionization is obtained from HEED, while all the transport properties are evaluated using MAGBOLTZ.
The transport dynamics have been followed using two different approaches.
In one, particle description using neBEM-Garfield++ combination has been used.
For this purpose, the neBEM solver has been significantly improved such that perturbations due to the charged particles present within the device are considered while estimating electric field.
In the other approach, the transport is simulated following hydrodynamic model using COMSOL during which the electric field is also provided by COMSOL where it is easy to set up space charge effects.
A comparison between these possible approaches will be presented.
Effect of different simulation parameters will also be demonstrated using simple examples.
\end{abstract}


\end{frontmatter}

%\maketitle

%%\linenumbers

\section{Introduction}

Charging up is a phenomena commonly observed in gaseous detectors having dielectric materials exposed to the active volume of gas mixture where primaries and secondaries are generated, and electron multiplication occurs.
They can affect long-term stability of a detector and lead to response non-uniformity \cite{Vishal2021}.
Space charge accumulation occurs in gaseous detectors due to the presence of charged particles within the active gas volume before they are collected at suitable electrodes.
Existence of space charge can distort the applied electric field configuration, make detector response non-uniform and unstable, and has the potential to lead to discharges \cite{Prasant2021a}, affecting detector performance significantly.
Besides making a detector lose its stability in the immediate time scale, formation of discharges has the capability of accelerating detector aging and damaging it for good.
Thus, it is important to understand these phenomena using both experimental and numerical tools.
However, these topics, and the associated discharge formation process, are complex and vast.
As a result, it is difficult to build satisfactory numerical models for these phenomena because large number of charged particles, as well as different length and time scales, are involved.

In recent times, there have been a number of simulation attempts to improve the understanding of these processes.
For example, charging up for GEMs and Thick GEMs (ThGEM) have been studied in \cite{Correira2014} using the Garfield++ framework \cite{Garfield} (C++ version of Garfield) in conjunction with ANSYS \cite{Ansys} and COMSOL \cite{Comsol} as commercially available Finite Element Method (FEM) field solvers.
Similarly, space charge problems have been investigated in \cite{Prasant2021a, Sheharyar, Samuel, Jaydeep2021, Promita2023, Tanay2023}.
For these investigations, parallelization attempts have yielded significant advancements for particle models \cite{Sheharyar, Samuel, Tanay2023}.
Similarly, interesting developments in fluid models (initially proposed in \cite{Paulo} and extended in \cite{Filipo}) have been carried out in \cite{Prasant2021a, Jaydeep2021, Promita2023}.

In the present brief paper, an attempt will be made to only discuss recent developments of few numerical tools currently available to address these problems and their performance in some typical scenarios.
In particular, extension of existing neBEM \cite{neBEM} field-solver to improve particle-based models will be discussed.
Advancement of existing hydrodynamic models \cite{Prasant2021a, Jaydeep2021} to include effects of additional physics phenomena like charging up will also be touched upon.
A comparison between particle and hydrodynamic models will finally be carried out and likely future developments will be outlined.

The particle and fluid numerical models are discussed in section \ref{section:Models}.
The simulation implementation is detailed in section \ref{section:Implementation}.
The results obtained are described and analyzed in section \ref{section:ResDis}, followed by the concluding remarks in section \ref{section:Conclusion}.

\section{Numerical models}
\label{section:Models}

HEED \cite{Heed} has been used for primary-ionization calculations and Magboltz \cite{Magboltz} to estimate drift, diffusion as well as Townsend and attachment coefficients from within Garfield++.
Using primary ionization data and the knowledge of different transport, multiplication and attachment coefficients, both particle and fluid models have been used to model transport of charged particles within Gas Electron Multiplier (GEM) \cite{Sauli} and Resistive Plate Chamber (RPC) \cite{Santonico} detectors.
Recent developments in neBEM has allowed estimation of effects due to charging up and space charge accumulation from the perspective of a particle model.
COMSOL on the other hand, has been used to simulate the charged particle dynamics using a fluid description.
While space-charge effects are automatically included in the fluid model, charging up effects have additionally been included within the computation.

\subsection{neBEM improvements}
\begin{figure}[hbt]
	\centering
	\includegraphics[scale=0.45]{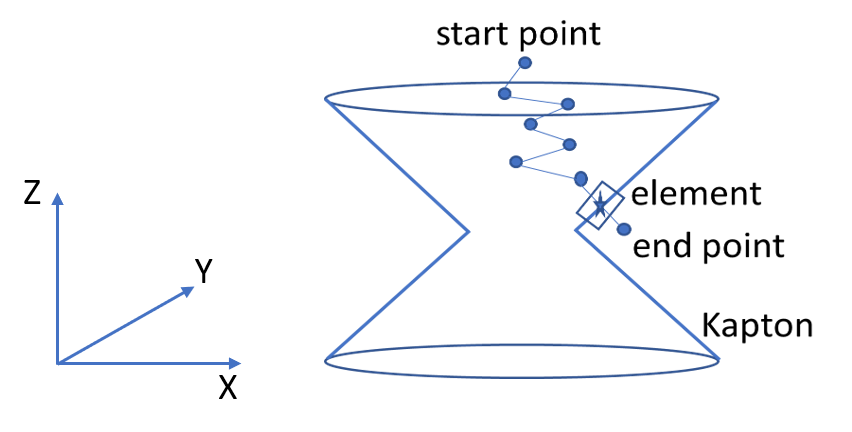}
	\caption{Identification of deposition location of a charged particle on the Kapton surface of a GEM hole.}
	\label{neBEMChargeDeposition}
\end{figure}

It may be mentioned here that neBEM discretizes any given geometry into a large number of small elements.
The initial device, set up using the Garfield++ interface, is represented as made up of a number of primitives (rectangles and right triangles) which, in their turn, are subdivided into small elements that are once again rectangle and right triangles.
When a charged particle gets deposited on an element that has dielectric properties, the charge gets attached to that surface for a long time, rather than being transported out of the detector by conducting electrodes. 
These locations can be obtained from Garfield++ and they can be directly used if we want to consider each charge individually.
In this case, the “end point” in the fig.\ref{neBEMChargeDeposition} will be the charge location.
However, if we have a large number of charged particles, this approach can be computationally very expensive.
A less demanding way can be to assign a surface charge density on the elements that collect the charges.
The way the element collecting a charge can be found has been indicated in figure \ref{neBEMChargeDeposition}.

Besides adding the capability of handling charging up effects, neBEM has also been improved to handle effect due to accumulation of charges within detector volume.
Once again, direct computation of influence of each charge is computationally expensive and different models have been implemented in the code to simplify computation.
By adopting a suitable model, it is possible to represent the charges as they really are (point charges), as uniformly charged lines, as uniformly charged areas and as volumes having equivalent point charges concentrated at the centroid.
It should be mentioned here that different versions of volume representation can lead to different Particle-In-Cell (PIC) algorithms.

\subsection{COMSOL improvements}
\begin{figure}[hbt]
	\centering
	\subfigure[]
	{\label{ComsolDevice}\includegraphics[scale=0.6]{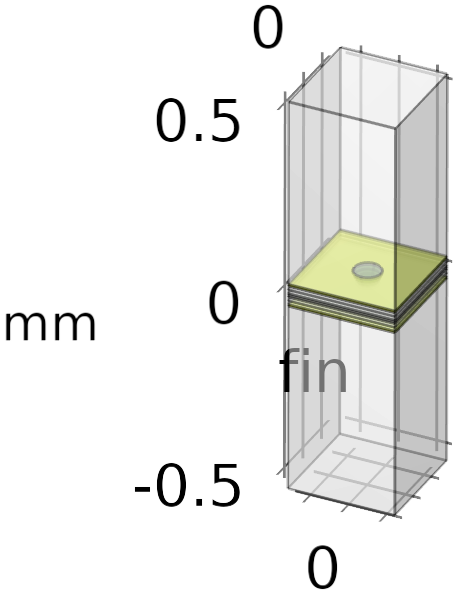}}
	\subfigure[]
	{\label{ComsolGEMIn}\includegraphics[scale=0.6]{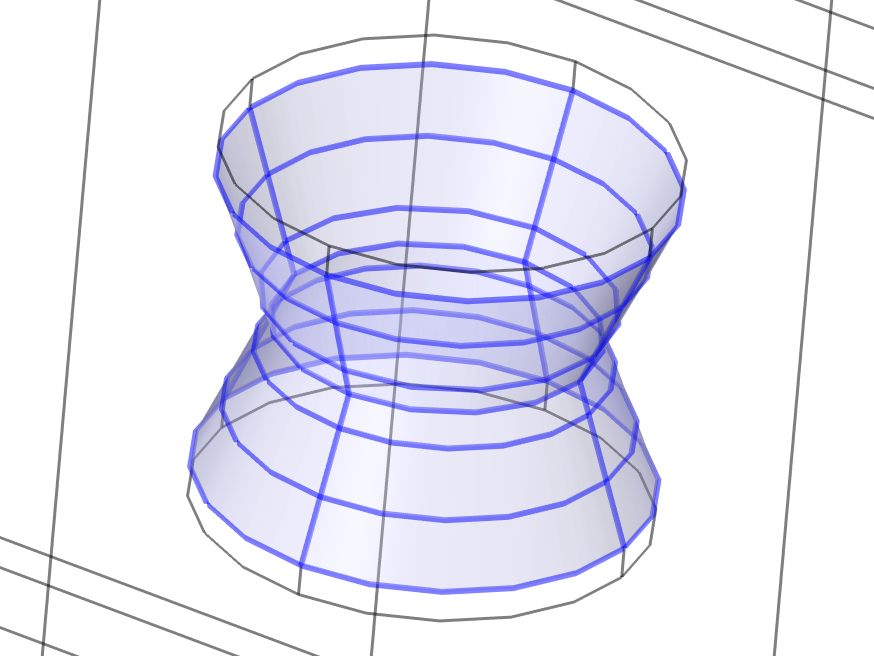}}
	\caption{A schematic view of the (a) GEM detector and (b) GEM hole inner surface in COMSOL.}
	\label{ComsolDetector}
\end{figure}

For solving transport of charged fluids, the device geometry is created using COMSOL, an example of which is shown in fig.\ref{ComsolDevice}.
The surface charge accumulation occurs on the Kapton surface exposed to the gas mixture, as shown in fig.\ref{ComsolGEMIn}.
The process is governed by the following equations

\begin{equation}
	\label{eqn:sca}
	\frac{\partial \rho_s}{\partial t} = \hat{n} . \vec{J_i} + \hat{n} . \vec{J_e} \\
\end{equation}
	
\begin{equation}
		-\hat{n} . (\vec{D_1} - \vec{D_2}) = \rho_s
\end{equation}
	
\noindent where $t$ is the time, $\hat{n}$ is the unit normal vector, $\rho_s$ is the surface charge density, $\vec{J_e}$ and $\vec{J_i}$ are the electronic and ionic current densities and $\vec{D_1}, \vec{D_2}$ are displacement currents.

As mentioned earlier, the fluid model automatically incorporates space-charge effects through the electrostatics Poisson equation 

\begin{equation}
	\label{eqn:es}
	\nabla . (\epsilon_0 \epsilon_r \vec{E}) = \rho_v
\end{equation}

\noindent where $\epsilon$ represents electrical permittivity, $\vec{E}$ is the electric field and $\rho_v$ is the space charge density.
For a given instant, the space charge density needs to be estimated throughout the detector volume as a function of space.
The equations governing the entire physics processes are well-described in \cite{Jaydeep2021} and not being repeated here.
It may noted here that for the presented results, ``physics governed" ``normal" mesh has been used in COMSOL.

\section{Implementation of numerical models}
\label{section:Implementation}

One single GEM-based detector and one RPC have been considered for the simulations presented in this paper.
The single GEM has 70-50-70 $\mu m$ biconical holes in copper clad Kapton foil of 50 $\mu m$.
The holes are arranged in the usual hexagonal pattern 140 $\mu m$ apart.
The RPC, on the other hand, has a single gas gap of 2mm.
It may be noted here that the gaseous mixture considered for all the results presented for GEM here is Ar-$\mathrm{CO_2}$ mixed in 70:30 ratio at atmospheric pressure, and for RPC it is 97\% Freon, 2.5\% isobutane and 0.5\% $\mathrm{SF_6}$.

\subsection{neBEM implementations}
\begin{figure}[hbt]
	\centering
	\subfigure[]
	{\label{eOnKapton1Hole}\includegraphics[scale=0.6]{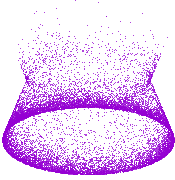}}
	\subfigure[]
	{\label{iOnKapton1Hole}\includegraphics[scale=0.6]{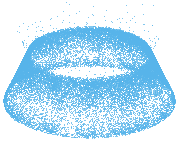}}
	\caption{Deposition of charged particles on the inner surfaces of a single hole in a GEM foil (a) electrons and (b) ions.}
	\label{ChargeDeposition1Hole}
\end{figure}

In order to observe the effect of radiation charging on an avalanche, we simulate the deposition of electrons and ions on Kapton surface of a GEM foil.
The pattern of charge deposition due to one single event indicates that while electrons are found in both halves of the bi-conical GEM hole, ions are more localized (figures \ref{eOnKapton1Hole} and \ref{iOnKapton1Hole}).
For this hole geometry, the ions are almost entirely found only in the GEM-half that faces the readout.
Number of ions close to the middle of the GEM hole is larger in comparison to electrons.
It is to be noted here that the pattern of charge deposition may change from event to event.
There are several parameters that can affect charge deposition including geometry of the hole, electric field configuration, nature of the material etc.
The case under consideration is one among many possibilities.

The ``end-point" locations have been used to compute charging up effects in one approach (termed as \textit{algo 1}), while in the other approach, charge deposited on each element has been computed (termed as \textit{algo 2}) to find out the effect.

Similarly, following the trajectory of each charged particle in Garfield++, it is possible to find out the space charge configuration at any given instant.
As indicated in section \ref{section:Models}, the locations of the charged particles can be used to set up a point, line, area or volume representation, as appropriate for a given problem.

\subsection{COMSOL implementations}
For surface charge accumulation at a given position on a dielectric material, the difference between electron and ion currents was computed (eq. \ref{eqn:sca}).
For space charge density $\rho_v$ in eq. \ref{eqn:es}, the difference between the number of electrons and ions per unit volume was considered.
The effects of surface charge density and space charge density were incorporated within the solution process by invoking multiphysics options, where necessary.
In order to simulate effects due to the accumulation of surface charge over a large number of events, charge density values obtained for one event were multiplied by 100.

\section{Results and discussions}
\label{section:ResDis}

\subsection{Charging up effects}
\subsubsection{Single event}
Here, we have compared $E_z$ and $E_x$ for a) surface charge represented by \textit{algo 1}, b) surface charge represented by \textit{algo 2}, and c) without any surface charge.
Since we are using only around $10^4$ charged particles for this computation, the effect of charging up is not pronounced in the $E_z$ component.
In $E_x$, there is perceptible difference among the values, although the overall magnitude in this case is small in comparison to $E_z$.
The maximum surface charge density, as estimated by \textit{algo 2}, on an element in the present case is around $10^{-7}$ $C/m^2$.
This small charge density, which is due to only one single event, is unable to have a pronounced effect on important parameters such as gain.
Please note that the coordinate system has been indicated in figure \ref{neBEMChargeDeposition}.

\begin{figure}[hbt]
	\centering
	\subfigure[]
	{\label{ChUpAxial}\includegraphics[scale=0.3]{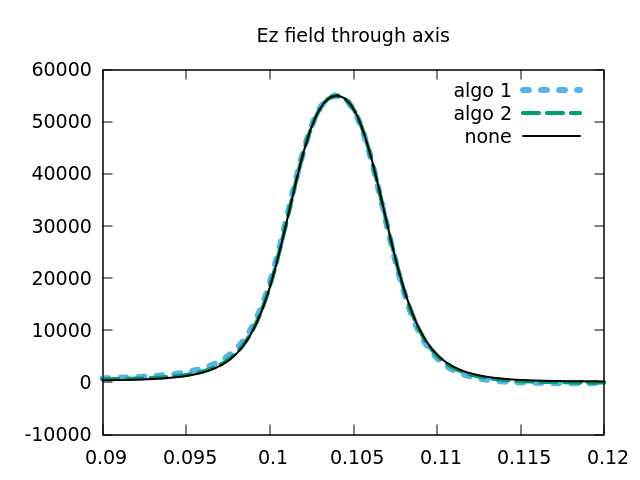}}
	\subfigure[]
	{\label{ChUpEx}\includegraphics[scale=0.3]{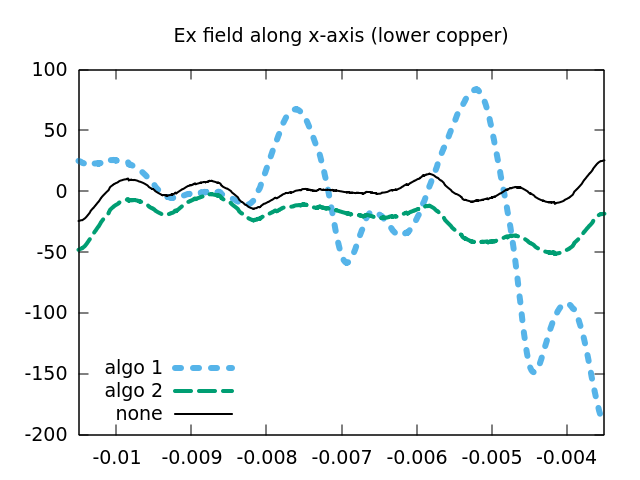}}
	\caption{Electric field variations due to charging up in a GEM-based detector using \textit{algo1}, \textit{algo2} and no charging up \textit{none}: (a) $E_z$ field along hole axis and (b) $E_x$ field variation along an X-line across GEM hole facing readout electrode.}
	\label{ChUpField}
\end{figure}

\begin{figure}[hbt]
	\centering
	\subfigure[]
	{\label{pointSCA}\includegraphics[scale=0.5]{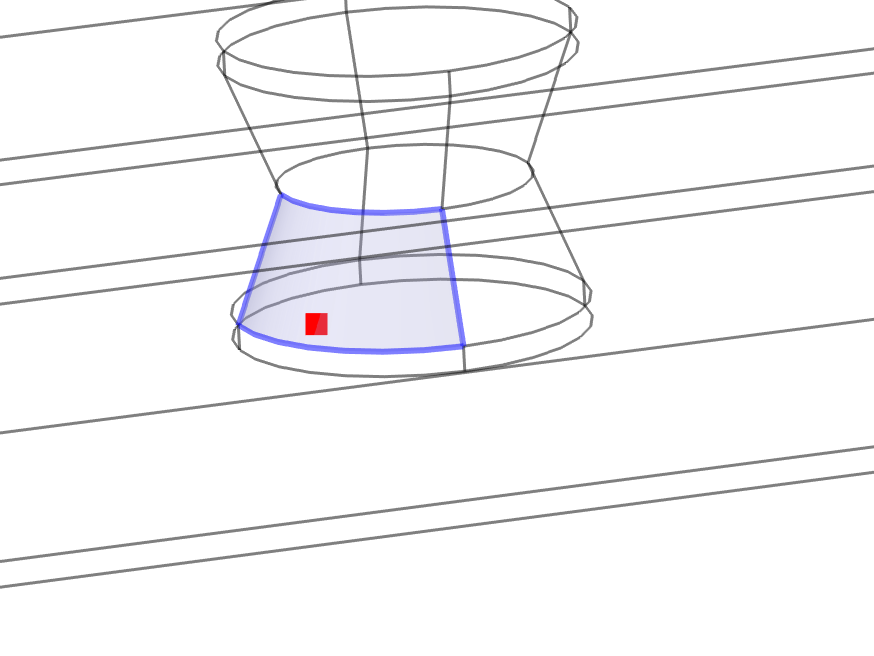}}
	\subfigure[]
	{\label{evolutionSCA}\includegraphics[scale=0.6]{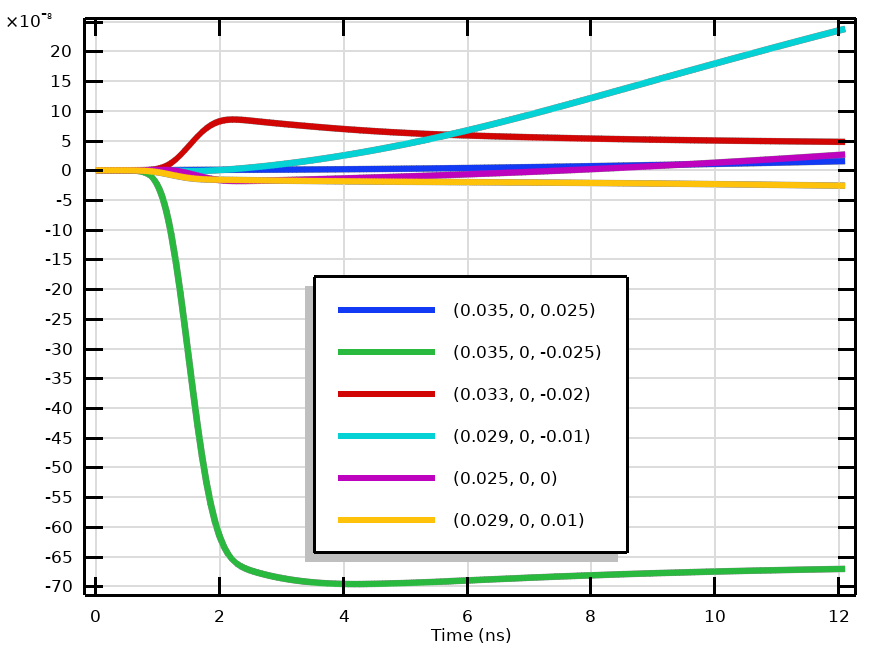}}
	\caption{Use of COMSOL to estimate charging up effects in a GEM-based detector: (a) A typical location on the Kapton inner surface where surface charge accumulation is estimated, and (b) surface charge accumulation over time at various locations on the Kapton inner surface.}
	\label{SCA}
\end{figure}

In order to cross-check the estimates of the particle approach, and also to evaluate the possibility of hydrodynamic modeling, we have simulated the dynamics of charge deposition on Kapton surface of a GEM hole for a similar event that was used in the particle model.
It may be noted that, due to the very nature of modeling mathematics, the events can only be made similar, but not identical.
The deposition of charge with time at various locations on the Kapton surface (one of the points is indicated as a red dot in fig.\ref{pointSCA}) is presented in fig.\ref{evolutionSCA}.

It can be seen that much negative charge deposition occurs close to the hole outlet facing the readout, as was also observed in the particle model. 
The maximum surface charge density is similar to that estimated by \textit{algo 2} in the particle model ($\approx$$50 \times 10^{-8}$  $C/m^2$) and it has been confirmed by further calculations that there is hardly any effect on gain.

\subsubsection{Charge accumulation over number of events}
Charges get deposited on Kapton surface for a large number of events before they start getting attached, or lost to a conductor.
In order to include the effects due to a large number of events, surface charge densities on Kapton surface equivalent to hundred times more than that estimated for a single event were specified.
Long-term charging up effects could, thus, be simulated in an approximate manner.
By this approach, keeping every other parameter unchanged, we could see that the gain has turned out to be as high as ~1000 in contrast to ~30 when no surface charge effects were considered. 
At this detector configuration, this large increase in gain seems to be indicative of the major effect that charging up has on detector response.
However, further detailed studies are necessary before it is possible to conclude on this topic.

\subsection{Space-charge effects}
\begin{figure}[hbt]
	\centering
	\includegraphics[scale=0.4]{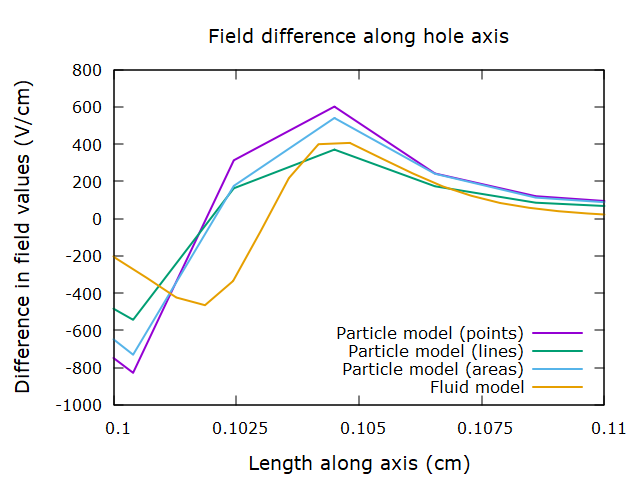}
	\caption{Estimation of space-charge effects in a GEM-based detector: Variation of $E_z$ along hole axis using various particle models and the fluid model.}
	\label{SpCh}
\end{figure}

Modification of electric field in a GEM-based detector has been estimated by using point, line and area representation of a given charged particle distribution.
Similar field modification has been estimated by using the COMSOL fluid model.
The field modifications estimated using different numerical models have been plotted in fig. \ref{SpCh}.

As can be observed, while there is significant quantitative mismatch, qualitatively the variation pattern is the same for all particle and fluid models.
The agreement between various particle models and the fluid model is particularly encouraging because the mathematical representation is quite different in these cases. 

\subsection{Formation of discharges}

\begin{figure}[hbt]
	\centering
	\subfigure[]
	{\label{RPCGain}\includegraphics[scale=0.25]{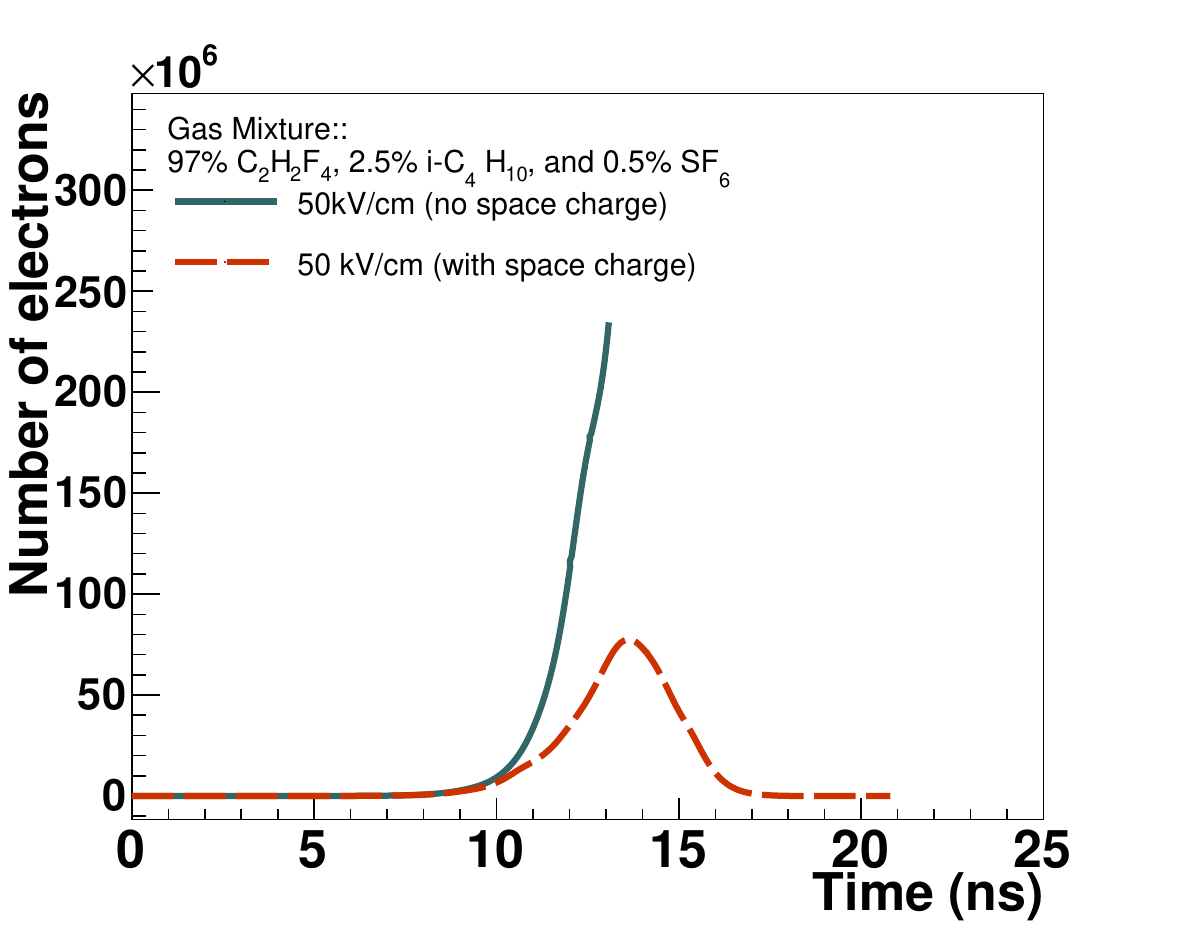}}
	\subfigure[]
	{\label{FluidStrForm}\includegraphics[scale=0.75]{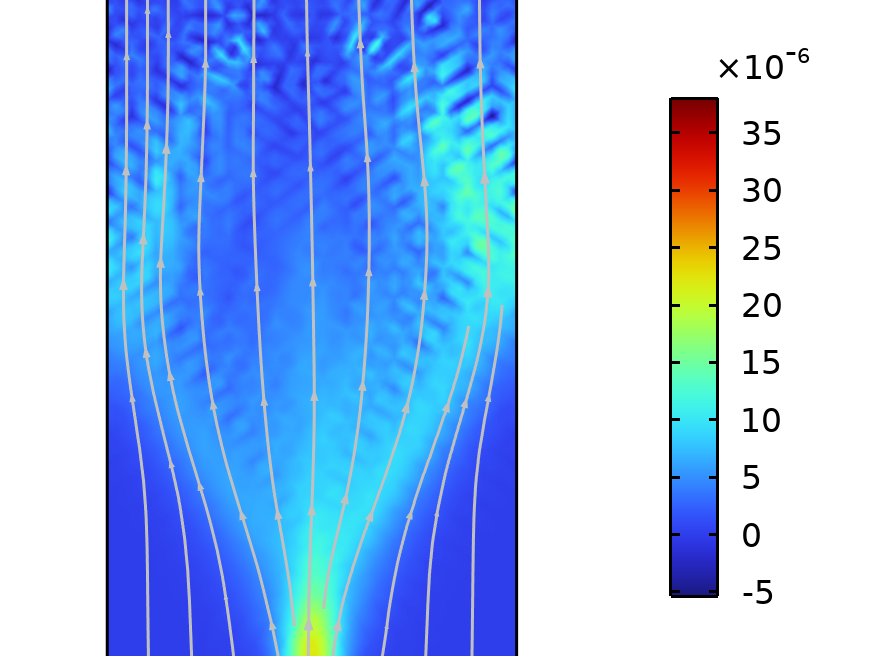}}
	\caption{(a) Effect of space charge on avalanche formation in an RPC at 50kV/cm using particle model, (b) ion density distribution during streamer formation in an RPC at 55kV/cm using fluid model.}
	\label{StrForm}
\end{figure}

Transition from avalanche to streamer formation in RPCs and GEMs has been studied using particle and fluid models.
OpenMP \cite{openmp} parallelization of neBEM and Garfield++ has been implemented (details in \cite{Tanay2023}) to carry out the work in particle model.
The strong effect of space charge in the estimation of transition from avalanche to streamer has been demonstrated in fig.\ref{RPCGain}.
In a particle model where the effect of space charge was ignored, the avalanche process seems to be unending, leading to the prediction of an unlikely streamer at an applied field of 50kV/cm.
However, when space-charge effect is included, the avalanche process is found to be well-contained.
It may be mentioned here that influence of negative ions on the electric field has also been found to be significant for these calculations \cite{Tanay2023}.
Example of a streamer occurring in an RPC at 55kV/cm has been separately simulated using fluid model that automatically includes space charge effects, as shown in fig.\ref{FluidStrForm}.

\begin{figure}[hbt]
	\centering
	\subfigure[]
	{\label{FieldStreamer}\includegraphics[scale=0.75]{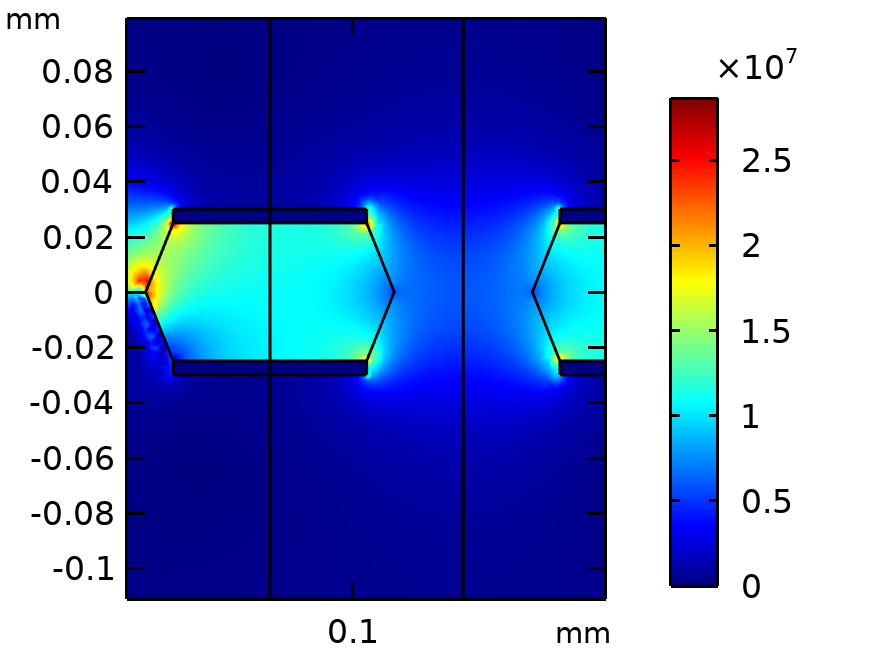}}
	\subfigure[]
	{\label{SingleGEMStrProb}\includegraphics[scale=0.3]{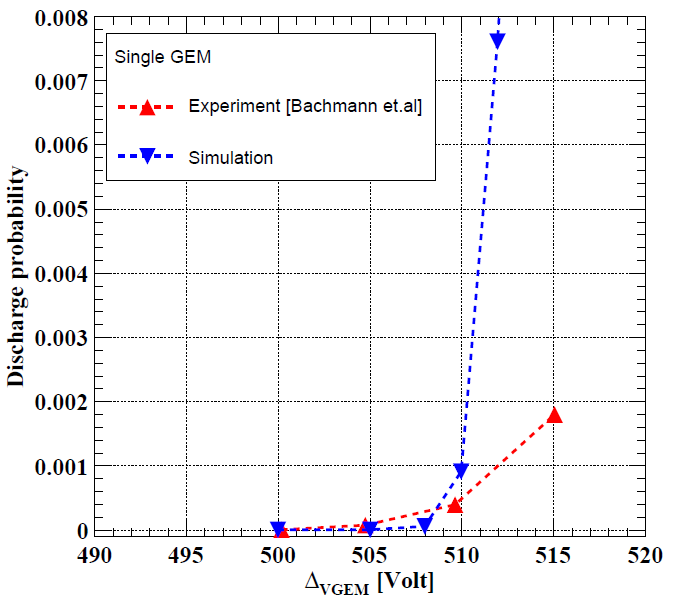}}
	\caption{(a) Field distortion leading to streamer formation in a GEM hole, (b) estimation of streamer probability in a single GEM-based detector.}
	\label{StrProb}
\end{figure}

Similarly, occurrence of discharge within a GEM hole has been simulated using the fluid model, as shown in fig. \ref{FieldStreamer}.
While the maximum applied field in this case was around 100kV/cm, the space charge effect increased the field to more than 250kV/cm, as shown in the figure.
By repeating the computation over a large number of similar configurations, discharge probability for a single GEM has been estimated using the fluid model (details in \cite{Prasant2021b}), as shown in fig. \ref{SingleGEMStrProb}.
Comparison with experimental results \cite{Bachmann} shows that the discharge probability estimates agree with the experimental values in a qualitative manner.

\section{Conclusion}
\label{section:Conclusion}

In summary, it can be said that several capable models have been developed to simulate phenomena related to charging up, space charge and discharge formation in a gaseous detector.
These developments are already producing qualitatively acceptable results.
As a result, it is possible to enrich our understanding of these phenomena using the numerical models.
However, they need much further refinement before being generally applicable.
For example, several possible representations are allowed for the particle model, but as yet there has been no attempt to optimize them for specific applications.
Moreover, accumulation of effects over a large number of events has not been attempted, except at an \textit{ad hoc} level.

Both particle and fluid approaches enjoy certain advantages and disadvantages over each other.
While particle models are very realistic and it is easy to incorporate statistical fluctuations, they are computationally extremely expensive and may need drastic simplifications to be applicable under usual circumstances.
Fluid models, on the other hand, are much less computationally demanding.
However, they are usually less realistic and it may be difficult to include statistical fluctuations.
Despite the inherent mathematical differences between the two approaches, predictions by both of them are found to be in general agreement, which is very encouraging.

\section{Acknowledgment}

We thank our respective Universities and Institutes for providing infrastructure necessary to carry out the research activities.
Part of this research has been performed in the framework of RD51 collaboration.
We wish to acknowledge the members of the RD51 collaboration for their help and suggestions, especially Dr. R. Veenhof and Dr. H. Schindler for their valuable comments.
This research was supported in part by the SERB research grant SRG/2022/000531. Author Dr. P. Bhattacharya acknowledges the Science and Engineering Research Board and SRG scheme for the necessary support.

\end{document}